# Nonreciprocal vortex isolator by stimulated Brillouin scattering in chiral photonic crystal fibre

Xinglin Zeng[1], Philip St.J. Russell[1], Christian Wolff[3], Michael H. Frosz[1], Gordon K. L. Wong[1] and Birgit Stiller[1,2]

[1]Max Planck Institute for the Science of Light, Staudtstr. 2, 91058 Erlangen, Germany
[2]Department of Physics, Friedrich-Alexander University, Staudtstr. 2, 91058 Erlangen, Germany
[3]Center for Nano Optics, University of Southern Denmark, Campusvej 55, DK-5230 Odense M, Denmark

*Corresponding author: xinglin.zeng@mpl.mpg.de

**Optical non-reciprocity, which breaks the symmetry between forward and backward propagating optical waves, has become vital in photonic systems and enables many key devices, such as optical isolators, circulators and optical routers. Most conventional optical isolators involve magneto-optic materials, but devices based on optical nonlinearities, optomechanically induced transparency and stimulated Brillouin scattering (SBS) have also been demonstrated. So far, however, they have only been implemented for linearly or randomly polarized $LP_{01}$-like fundamental modes. Here we report a light-driven nonreciprocal isolator for optical vortex modes, based on topology-selective SBS in chiral photonic crystal fibre. The device can be reconfigured as an amplifier or an isolator by adjusting the frequency of the control signal. The experimental results show vortex isolation of 22 dB, which is at the state-of-the-art in fundamental mode isolators using SBS. This unique device may find applications in optical communications, fibre lasers, quantum information processing and optical tweezers.**

Breaking two-way optical symmetry so as to achieve non-reciprocal isolation is crucial in many laser, amplifier, optical communications and sensing systems [1]. In particular it is of great importance in all-optical signal routing, protecting lasers from disruptive back-reflections, and reducing multipath interference in optical communications. Moreover, isolators generally improve the performance of optical systems through suppression of unwanted interference, inter-device interactions and channel cross-talk. Most conventional optical isolators are based on the Faraday effect in magneto-optical materials [2,3]. Alternative approaches, such as optical isolation through the Kerr effect [4,5], optomechanically induced transparency [6,7], spatial-temporal modulation [8,9] and stimulated Brillouin scattering (SBS) [10-12], have also been proposed and demonstrated, though only for linearly or randomly polarized $LP_{01}$-like fundamental modes.

Circularly polarized vortex beams or modes have been extensively studied in recent years, in connection with applications in quantum and classical communications [13,14], optical tweezers [15] and quantum information processing [16]. As a result, devices such as vortex generators, lasers and signal amplifiers have been demonstrated and are in great demand [17]. A device that is so far missing is a vortex isolator, which is essential for high-power vortex lasers and optical communications.

Here we report the first example of a reconfigurable light-driven non-reciprocal isolator for circularly polarized vortex modes, based on topology-selective stimulated Brillouin scattering (SBS) in chiral photonic crystal fibre (PCF). In recent years chiral photonic crystal fibre [18] − which offers a unique platform for studying the behaviour of light in chiral structures that are infinitely extended in the direction of the twist − has been shown to robustly preserve optical modes carrying circular polarization states and optical vortices over long distances, allowing investigation of nonlinear processes in the presence of chirality [19,20]. We report isolation of vortex modes through topology-selective SBS in chiral PCFs with three-fold-rotational symmetry (C3 PCF) and six-fold-rotational symmetry (C6 PCF). In particular, angular momentum conservation dictates that the topological charge and spin of the backward Brillouin signal are opposite to those of the pump. The experiments show isolation factors higher than 22 dB for two different vortex orders, which is comparable with the best SBS-related optical isolators for fundamental modes [21]. The isolation remains nearly constant within 2 dB over a 35 dB dynamic range of signal power, indicating excellent optical linearity. Switching between nonreciprocal isolation and amplification can be simply achieved by down- or up-shifting the control signal frequency by the Brillouin frequency. We also develop an analytical theory for the dynamics of nonreciprocal SBS in chiral PCF and achieve good agreement with experimental measurements.

**Results**

**Topology-selective SBS in chiral PCF.** $N$-fold rotationally symmetrical (symmetry class $C_N$) chiral PCFs support helical Bloch modes (HBMs) [22], whose $m$-th order azimuthal harmonics carry optical vortices with azimuthal order $\ell_A^{(m)} = \ell_A^{(0)} + Nm$, where $\ell_A^{(m)}$ is the number of complete periods of phase progression around the azimuth for fields expressed in cylindrical components and the $\ell_A^{(0)}$ is the principal azimuthal order. Note that $\ell_A$ is always an integer and is robustly conserved. In chiral PCF it is found, both experimentally and by numerical modelling, that the fields are almost perfectly circularly polarized. In this case the cylindrical transverse electric fields can be expressed in Cartesian components simply by multiplying them with a rotation matrix, leading to an expression that links the azimuthal order $\ell_A^{(m)}$ to the topological charge $\ell_T^{(m)}$: $\ell_A^{(m)} = \ell_T^{(m)} + s$, where spin $s = +1$ denotes left circular polarization state. Here, we use the shorthand [$\ell_T$, $s$] to denote the parameters of the circularly polarized vortex-carrying HBMs, where for ease of notation $\ell_T = \ell_T^{(0)}$ is defined as the principal topological order. In chiral PCF, HBMs with equal and opposite values of $\ell_T$ are generally non-degenerate in index, i.e., topologically birefringent, while modes with opposite spin but the same value of $\ell_T$ are weakly birefringent [22].

Topology-selective SBS in chiral PCF is a process in which a forward pump (P) mode scatters strongly into a backward Stokes (S) mode when [$\ell_T$, $s$]$_P$ = –[$\ell_T$, $s$]$_S$, but shows no interaction when [$\ell_T$, $s$]$_P$ = [$\ell_T$, $s$]$_S$, as illustrated in Fig. 1a. The fields of the forward (pump) and backward (Stokes) HBMs can be written as:

$$\mathbf{E}_P(\mathbf{r},t) = A_P(z)a(x,y)\mathbf{u}_P\, e^{i(\beta_P z + \ell_{TP}\phi)}e^{-i\omega_P t}$$
$$\mathbf{E}_S(\mathbf{r},t) = A_S(z)a(x,y)\mathbf{u}_S^*\, e^{-i(\beta_S z + \ell_{TS}\phi)}e^{-i\omega_S t} \qquad (1)$$

where $\mathbf{u}_k = (u_{kx}, u_{ky}) = (1, is_k)/\sqrt{2}$ is a complex unit vector ($\mathbf{u}_k \cdot \mathbf{u}_k^* = 1$) representing the polarization state, $\ell_{Tk}$ is the topological charge, subscript $k = P$ or $S$ denotes pump or Stokes, $A_k(z)$ is a slowly varying amplitude, $a(x,y)$ the scalar modal field distribution (assumed identical for both modes), $\beta_k$ the propagation constant and $\omega_k/2\pi$ the optical frequency. The vortex-free acoustic density wave excited by Brillouin scattering may similarly be written:

$$\rho(\mathbf{r},t) = \rho_m a_q(x,y)\, e^{i(qz - \Omega t)} \qquad (2)$$

where $q$ and $\Omega/2\pi$ are the acoustic wavevector and frequency. The optoacoustic overlap is then proportional to the integral over the transverse cross-section of:

$$(\mathbf{E}_P \cdot \mathbf{E}_S^*)\rho^* = A_P A_S^* \rho^* a^2 a_q\, e^{i(\ell_{TP} + \ell_{TS})\phi}(\mathbf{u}_P \cdot \mathbf{u}_S)e^{i(\beta_P + \beta_S - q)z - i(\omega_P - \omega_S - \Omega)t} \qquad (3)$$

which is non-zero only if $\ell_{TP} = -\ell_{TS}$ and $\mathbf{u}_P \cdot \mathbf{u}_S = 1$, i.e., $s_P = -s_S$. The Brillouin gain is maximized when both energy and momentum are conserved, i.e., $\Omega = \omega_P - \omega_S$ and $q = \beta_P - \beta_S$.

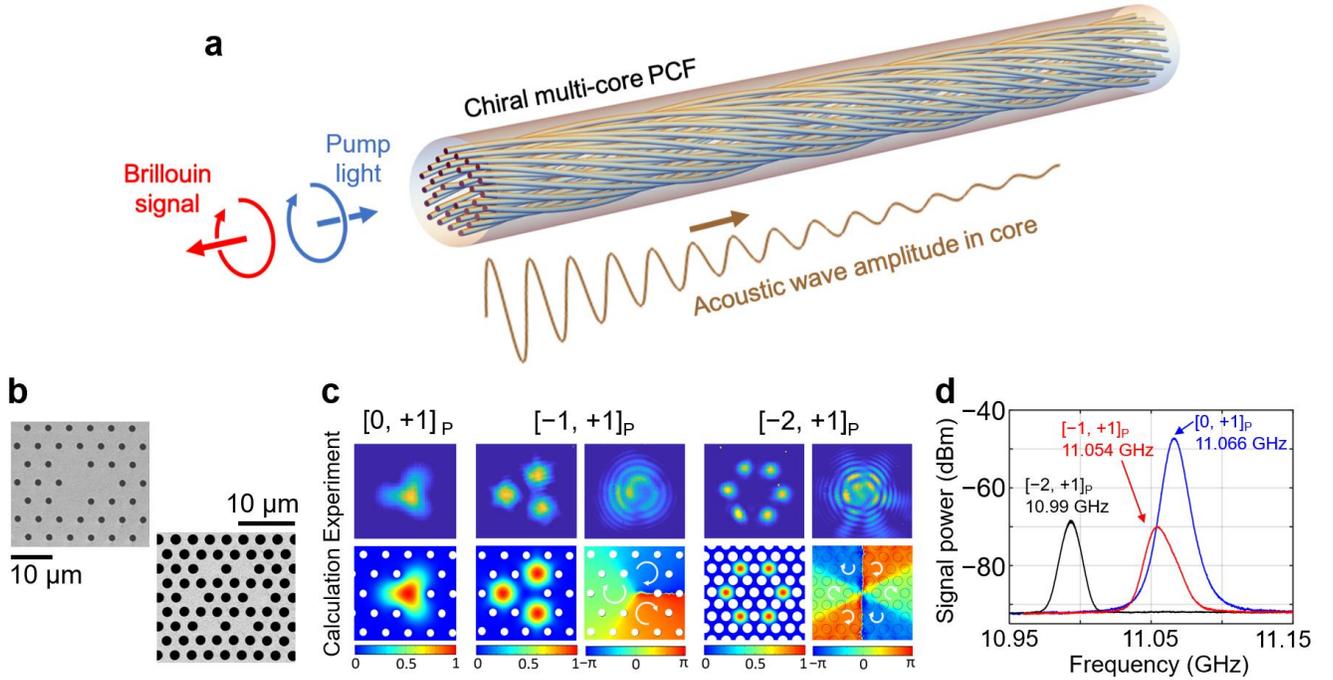

**Fig. 1 | Topology-selective Brillouin scattering in chiral PCF. a,** Conceptual view of SBS in chiral three-fold rotationally symmetric PCF. The circular arrows indicate both the spin and the direction of azimuthal phase progression, which are preserved during SBS. As a result, pump and Stokes signals interact only when the signs of $\ell_T$ and $s$ are reversed relative to the beam directions. **b,** Scanning electron micrographs of the $C_3$ PCF and $C_6$ PCF used in the experiments. **c,** Experimentally measured and numerically calculated near-field distributions for $[\ell_T, s]_P = [0, +1]$ and $[-1, +1]$ in $C_3$ PCF and $[\ell_T, s]_P = [-2, +1]$ in $C_6$ PCF. The subscript P denotes pump wave. Interference patterns between the vortex-carrying HBMs and a divergent Gaussian beam are shown in the upper-right-hand panel of each higher-order HBM. The calculated polarization and phase distributions are shown in the lower-right-hand panel of each higher-order HBM. The measured loss of $[0, \pm1]$ and $[\pm1, \pm1]$ modes in $C_3$ PCF are 0.013 dB/m and 0.017 dB/m and the loss of the $[\pm2, \pm1]$ modes in $C_6$ PCF are 0.04 dB/m. The index difference is $\sim 8 \times 10^{-4}$ between $[+1, \pm1]$ and $[-1, \pm1]$ modes and $\sim 7 \times 10^{-4}$ between $[+2, \pm1]$ and $[-2, \pm1]$ modes, and the circular birefringence at fixed $\ell_T$ is $\sim 6 \times 10^{-6}$ in $C_3$ PCF and $\sim 2 \times 10^{-5}$ in $C_6$ PCF. **d,** Spontaneous Brillouin spectra generated by pumping $C_3$ PCF (at 1550 nm) with $[0, +1]_P$ (blue) and $[-1, +1]_P$ (red) modes, and $C_6$ PCF with the $[-2, +1]_P$ (black) mode.

**Noise-initiated SBS measurement.** Two different chiral PCFs were used in the experiments. The first had 3-fold rotational symmetry ($C_3$ PCF) and a twist period of 5 mm [23] and the second 6-fold rotational symmetry ($C_6$ PCF) and a twist period of 7.2 mm. The preforms were constructed by the standard stack-and-draw process and the fibres drawn from a spinning preform. Scanning electron micrographs (SEMs) of the two PCFs are shown in Fig. 1b. The hollow channel diameter $d$ was 1.6 µm and inter-channel spacing $\Lambda$ was 5.16 µm for the $C_3$ PCF, and $d = 2$ µm, $\Lambda = 2.98$ µm for the $C_6$ PCF. Fig. 1c shows measured near-field intensity profiles, alongside numerical simulations, for $[0, +1]_P$ and $[-1, +1]_P$ HBMs after propagation along 200 m of $C_3$ PCF, and the $[-2, +1]_P$ HBM after propagation along 200 m of $C_6$ PCF. For each vortex mode, the upper right-hand panel shows the spiral fringe pattern formed by interference with a divergent Gaussian beam; the lower panels are the corresponding calculated phase distributions. The modulus of the measured Stokes parameter, $|S_3|$, is higher than 0.98 at the output of both PCFs, showing very good preservation of spin as well as topological charge, even after 200 m of propagation.

Fig. 1d shows a heterodyne measurement of the backscattered Stokes signal, formed by mixing with a local oscillator (LO) to produce an electrical beat-note in the GHz domain (for details see Supplementary Section S1). The Brillouin frequency shifts were 11.066 GHz for $[0, \pm1]_P$ and 11.054 GHz for $[\pm1, \pm1]_P$ in $C_3$ PCF, and 10.99 GHz for $[\pm2, \pm1]_P$ in $C_6$ PCF. Numerical calculations using finite element modelling (FEM) predict corresponding frequencies of 11.085 GHz, 11.063 GHz and 11.027 GHz, in good agreement with experiment. Compared to PCFs with high air-filling fractions and μm-scale cores, which guide several different types of acoustic core modes and produce a multi-peaked Brillouin spectrum [24], the relatively low air-filling fraction and large core diameter of the two chiral PCFs produces a Brillouin frequency shift close to that of bulk glass: $2n_s v_L/\lambda_p = 11.12$ GHz, where $n_s$ is the refractive index of silica at the pump laser wavelength $\lambda_p$, and $v_L = 5971$ m/s is the longitudinal acoustic phase velocity.

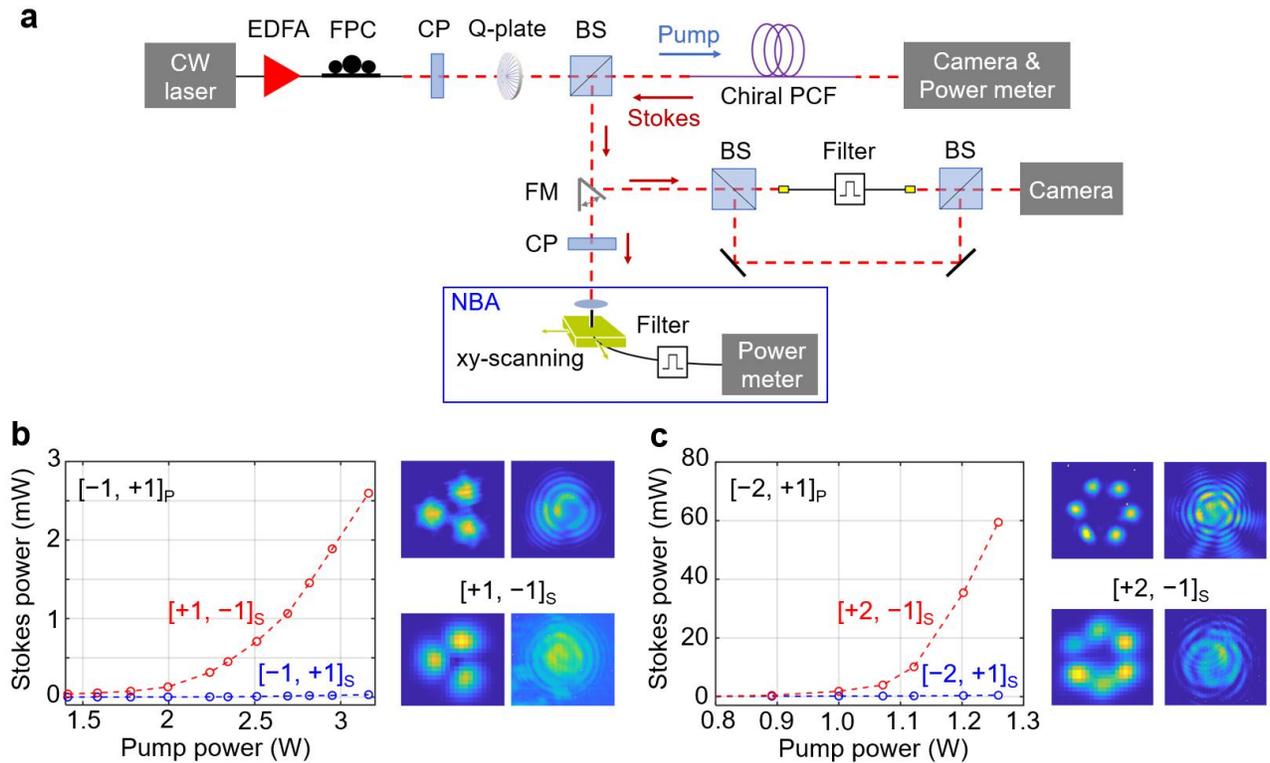

**Fig. 2 | Noise-initiated topology-selective SBS in chiral PCF. a,** Experimental setup for measuring the Stokes signal with vortex pumping. FPC, fibre polarization controller; CP, circular polarizer; BS, beam splitter; FM, flip-mirror; Q-plate, an optical device that can generate vortex-carrying light beams from a circularly polarized Gaussian beam; NBA: near-field Brillouin scanning analyser. **b,** Stokes signal power, polarization state, mode profile and topological charge for pumping $C_3$ PCF with a $[-1, +1]_P$ mode. **c,** Same as **b**, for pumping $C_6$ PCF with a $[-2, +1]_P$ mode. The broad bandwidth (~40 MHz) of the spontaneous Stokes signals somewhat obscures the Stokes interference patterns, while the less noisy pump patterns are much cleaner.

We next increased the pump power above threshold so as to reach the SBS regime, when the polarization states, mode profiles and topological charges of the much stronger Stokes signals could be more easily and precisely measured. Only the measurements for vortex-carrying ($[-1, +1]_P$ and $[-2, +1]_P$) HBM pumps are presented here (for completeness, measurements for $[0, \pm1]_P$ are available in Supplementary Section S2). Fig. 2a shows the experimental setup (for more details see "Methods"). Fig. 2b and 2c shows the power dependence of the Stokes signal for $[-1, +1]_P$ in $C_3$ PCF and $[-2, +1]_P$ in $C_6$ PCF. As already mentioned, the spin and topological charge of the Stokes signal are opposite in sign to the pump (see interferometric patterns in Fig. 2b and 2c). The robustness of spin and topological charge preservation in chiral PCF was further confirmed by measurement of the Stokes parameter $S_3$ of the Stokes signal for $[-1, +1]_P$ and $[-2, +1]_P$. The measured value of $|S_3|$ was greater than 0.96 in all cases (for details see Supplementary Section S3).

**SBS gain coefficient measurement.** We next measured the Brillouin gain coefficients $g_B$ for vortex modes in $C_3$ and $C_6$ PCF, both 200 m long. The amplified Stokes signal takes the well-known form [25]:

$$P_S(0) = P_S(L)\exp\left(g_B P_P (1-e^{-\alpha L})/\alpha - \alpha L\right) \tag{4}$$

where $\alpha$ is the fibre loss (0.017 dB/m for $[\pm 1, \pm 1]$ modes in $C_3$ PCF and 0.04 dB/m for $[\pm 2, \pm 1]$ modes in $C_6$ PCF), $L$ the fibre length and $P_P$ the pump power. Details of the experimental setup are available in Supplementary Section S4. Figure 3 plots $g_B$, estimated from the experimental measurements using Eq.(4), as a function of pump-seed frequency difference at a pump power of 0.8 W and a Stokes seed power of 10 mW. As expected, the gain is significant only when pump and seed have opposite topological charge and spin, reaching peak values of 0.022 m$^{-1}$W$^{-1}$ for $[-1, +1]_P$ and $[+1, -1]_S$ in $C_3$ PCF and 0.185 m$^{-1}$W$^{-1}$ for $[-2, +1]_P$ and $[+2, -1]_S$ in $C_6$ PCF. These two measured peak gain coefficients are close to the theoretical values of 0.032 m$^{-1}$W$^{-1}$ and 0.162 m$^{-1}$W$^{-1}$, calculated using the theoretical expression for the line-centre gain coefficient, taking account of the optoacoustic overlap [26]. Note that since the topological charge of the Stokes signal is opposite in sign to the pump, the identical acoustic mode is excited for both signs of $\ell_{TP}$. For completeness, the measured gain spectra for $[0, +1]_P$ and $[0, -1]_S$ modes in $C_3$ PCF are available in Supplementary Section S4, showing peak gain coefficients of 0.169 m$^{-1}$W$^{-1}$.

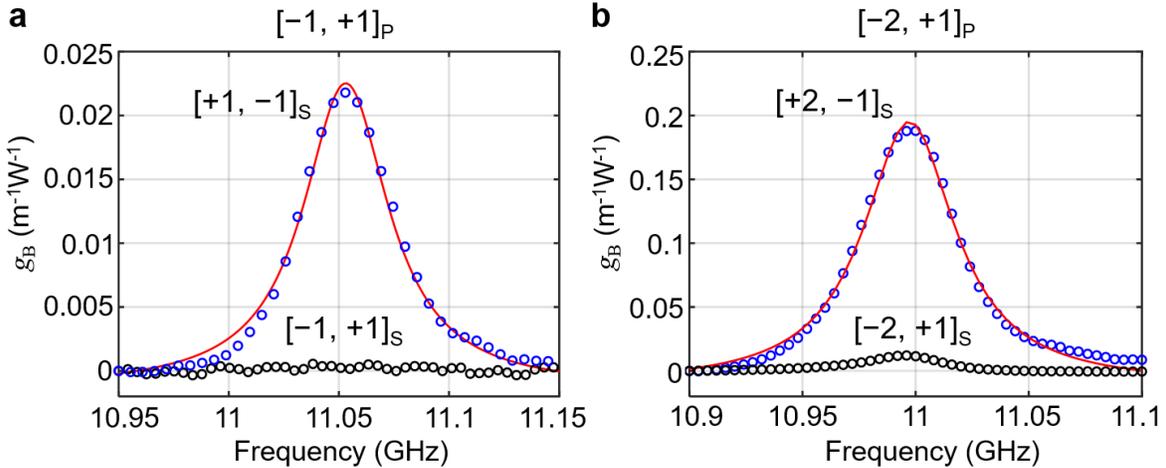

**Fig. 3 | Brillouin gain measurement in chiral PCF. a,** Frequency dependence of the Brillouin gain $g_B$ in $C_3$ PCF pumped by a $[-1, +1]_P$ mode. The circles are measured data and the red lines are Lorentzian fits. **b,** Same as **a** but for $C_6$ PCF pumped by the $[-2, 1]_P$ mode.

**Reconfigurable light-driven vortex isolator.** Next we used the setup to demonstrate a nonreciprocal vortex isolator (Fig. 4). Each circularly polarized vortex-carrying beam passes through two λ/4 plates placed on opposite sides of a polarizing beam-splitter (PBS), permitting the polarization state in each path to be sequentially converted from circular to linear and from linear to circular. Where necessary, we denote control parameters with the subscript "ctrl" and signal parameters by subscript "sig". When a signal (parameters $[\ell_T, s]_{sig}$) at frequency $f_0$ is launched into the fibre, along with a counter-propagating control wave (parameters $[-\ell_T, -s]_{ctrl}$) at frequency $f_0 - f_{SBS}$, power is transferred from the signal to the control wave and the signal is attenuated. In contrast, a counter-propagating signal with parameters $[\ell_T, s]_{sig}$ at frequency $f_0$ is unaffected. If the control wave frequency is changed to $f_0 + f_{SBS}$, keeping all other parameters constant, power is transferred from the control to the signal, which is amplified. Once again, a counter-propagating signal (parameters $[\ell_T, s]_{sig}$) at frequency $f_0$ is unaffected. The system can thus be conveniently configured as a nonreciprocal amplifier or attenuator by adjusting the control signal frequency.

Figure 5a shows the experimental setup for observing nonreciprocity (see "Methods" for more details). Figure 5b tabulates the measured power in the transmitted forward and backward signals in each case, for example, a forward $[-1, +1]_{sig}$ mode is strongly attenuated by a backward $[+1, -1]_{ctrl}$ mode in $C_3$ PCF, while the backward $[-1, +1]_{sig}$ mode is unaffected. The dependence on control power of the attenuation and amplification of $[-1, +1]_{sig}$ and $[-2, +1]_{sig}$ modes is explored in Fig. 6a and Fig. 6b. For a signal power of 617 mW, the isolation is 22.2 dB for the $[-1, +1]_{sig}$ mode (left figure, blue) and 23.4 dB for the $[-2, +1]_{sig}$ mode (right figure, blue). By switching the frequency of the control signal from $f_0 - f_{SBS}$ to $f_0 + f_{SBS}$, nonreciprocal amplification factors of 21 dB for the $[-1, +1]_{sig}$ mode (left-hand figure, red) and 18 dB for the $[-2, +1]_{sig}$ mode (right-hand figure, red) were obtained, keeping the signal power at 12.6 mW.

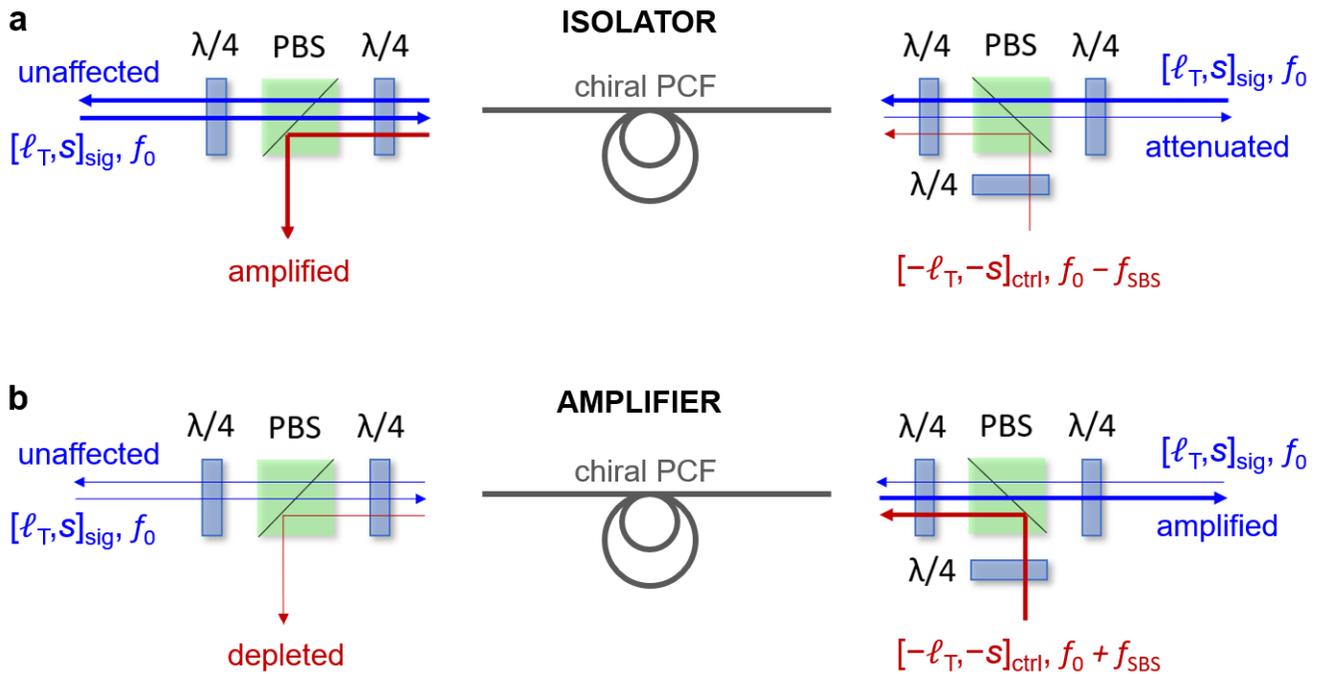

**Fig. 4 | Schematic diagrams of isolator and amplifier. a**, Nonreciprocal optical isolator. The control signal is combined with, or separated from, the signal using polarizing beam-splitters (PBS). When a backward control wave $[-\ell_T, -s]_{ctrl}$ with frequency $f_0 - f_{SBS}$ is launched, the forward signal $[\ell_T, s]_{sig}$ is highly attenuated while the backward-propagating signal $[\ell_T, s]_{sig}$ is unaffected. **b**, When the backward control wave frequency is changed to $f_0 + f_{SBS}$, the forward signal $[\ell_T, s]_{sig}$ is amplified while the backward-propagating signal $[\ell_T, s]_{sig}$ is unaffected.

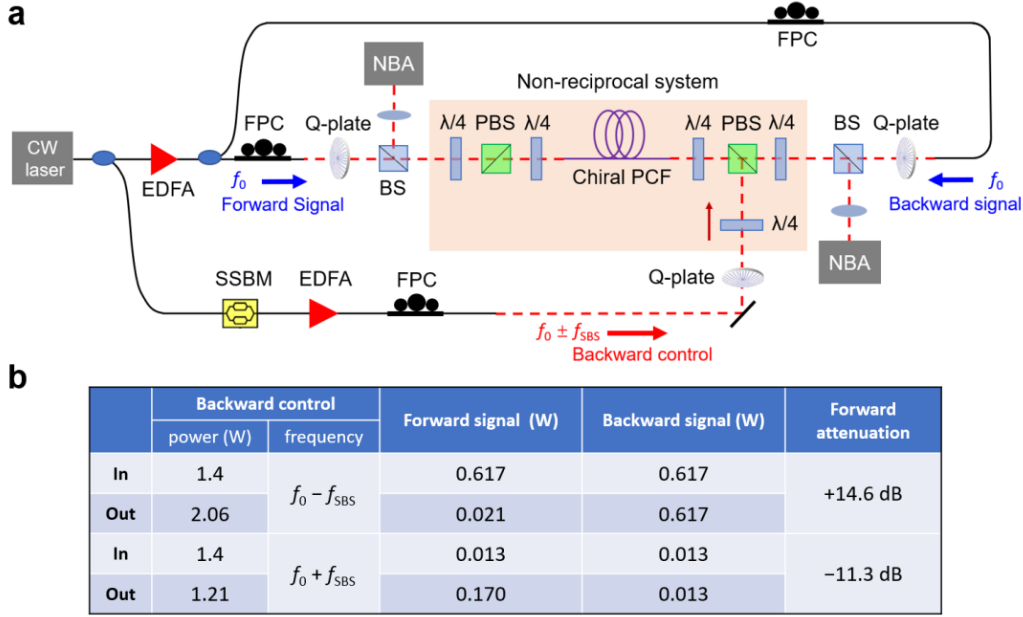

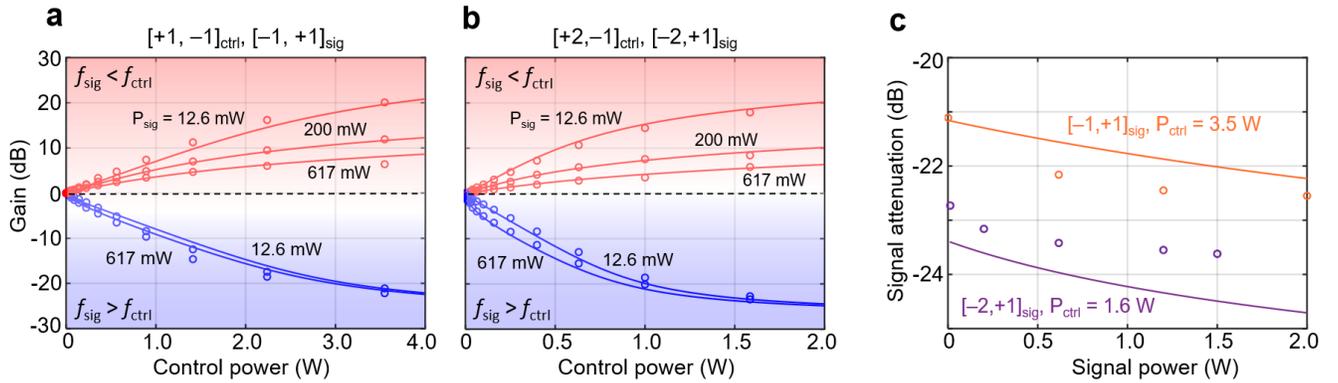

**Fig. 5 | Experimental setup and measurements. a**, Schematic diagram of the reconfigurable light-driven vortex isolator. The chiral fibre was 200 m long in all the experiments. **b**, Nonreciprocal isolation and amplification measurements for [−1, +1] modes. The forward signal waves are strongly depleted (upper two rows) or amplified (lower two rows) by backward control waves, while the backward signal wave is unaffected.

**Fig. 6 | Nonreciprocal attenuation and amplification. a**, Dependence of $[-1, +1]_{sig}$ gain on $[+1, -1]_{ctrl}$ power for $f_{sig} < f_{ctrl}$ (upper red-shaded region) and $f_{sig} > f_{ctrl}$ (lower blue-shaded region). $P_{sig}$: signal power, $P_{ctrl}$: control power. **b**, Same as **a** but for $[-2, +1]_{sig}$ and $[+2, -1]_{ctrl}$ modes. **c**, Dependence of signal attenuation on signal power. The circles are experimental data points and the curves theoretical predictions based on Eq. 5.

It is notable that the isolation saturates with increasing control light power (Fig. 6a and 6b). We attribute this to the onset of second order SBS when the first Stokes signal becomes strong. This is aided by second harmonic distortion in the single-sideband modulator (SSBM), which generates a signal at ($f_0 − 2f_{SBS}$) that is ~27 dB weaker than the first harmonic. This signal is then Fresnel-reflected at the input face of the fibre, thereby seeding the second Stokes signal. The well-known coupled wave theory of SBS, based on slowly varying amplitudes, can be used to treat this case [27], leading to the coupled power equations:

$$\begin{aligned}
\frac{\partial P_P}{\partial z} &= -(\alpha + g_B P_{S1}) P_P \\
\frac{\partial P_{S1}}{\partial z} &= (\alpha + g_B (P_{S2} - P_P)) P_{S1} \\
\frac{\partial P_{S2}}{\partial z} &= (-\alpha + g_B P_{S1}) P_{S2}
\end{aligned} \qquad (5)$$

where $P_P$ is the pump power, $P_{S1}$ the first (backward) Stokes power and $P_{S2}$ is the second (forward) Stokes power. The exponential power attenuation rate is $\alpha$ m$^{-1}$ and $g_B$ is the effective Brillouin gain in units of m$^{-1}$W$^{-1}$. Experimental estimates of $g_B$ are 0.022 m$^{-1}$W$^{-1}$ for the C$_3$ PCF and 0.185 m$^{-1}$W$^{-1}$ for the C$_6$ PCF. Although the Brillouin frequency shifts for P-to-S1 and S1-to-S2 conversion are not strictly the same, the difference is only 1 part in $10^4$, so can be neglected. As shown in Fig. 6a and 6b, numerical solutions of Eq. (5) are in good agreement with the measurements, confirming that the saturation is caused by cascaded SBS.

In applications such as optical communications and all-optical signal processing, it is vital that isolation is effective over wide range of different signal powers. The chiral fibre performs very well, offering ~22 dB isolation within ±1 dB for the [−1, +1]$_{sig}$ mode (orange circles in Fig. 6c) over a 35 dB range of signal power at a control power of 3.55 W. The same is true for the [−2, +1]$_{sig}$ mode (purple circles in Fig. 6c) at a control power of 1.59 W. The measurements are in reasonable agreement with theoretical calculations, as shown by solid curves in Fig. 6c.

**Discussions and conclusions**

Efficient light-driven optical isolators for circularly polarized HBMs can be realised by topology-selective stimulated Brillouin scattering in chiral $N$-fold rotationally symmetric multi-core PCF. HBMs with principal topological charge up to the nearest integer below $|N/2|$ can be guided and isolated. Non-reciprocal behaviour follows because Brillouin gain only exists when the Stokes and pump signals have equal and opposite topological charge and spin. Isolation factors greater than 22 dB are obtained over a 35 dB dynamic range of input signal. Non-reciprocal vortex amplification can be achieved by using a control signal with frequency $f_{SBS}$ above the vortex signal. Although the working bandwidth is limited by the Brillouin gain linewidth, isolation over a much wider bandwidth is possible if the control wavelength is tuned in synchronism with the signal wavelength.

Finally, we mention that the Brillouin gain would be some 100 times higher if non-silica glasses such as chalcogenides As$_2$Se$_3$ or As$_2$S$_3$ [28] were used instead of silica. Although such fibres are challenging to fabricate, they would permit efficient vortex isolation at much shorter [~1 m] fibre lengths. In addition, if implemented using short control pulses, light-driven vortex isolators could be useful in many different all-optical systems [10], as well as in vortex lasers that are employed in optical trapping, communications and quantum entanglement.

**Acknowledgement:** We acknowledge support from the Max-Planck-Gesellschaft. We thank Yang Chen and Zheqi Wang for help with several aspects of theory and experiments.

**Competing interests:** Authors declare no competing interests.

**Author contributions:** The concept was proposed by X.Z., P.St.J.R. and B.S., X.Z. performed the experiments. X.Z. and C.W. performed the simulation and theoretical calculation. M.F. and G.W. fabricated the chiral PCF. The results were analysed by X.Z., P.St.J.R. and B.S. and the manuscript was written by X.Z., P.St.J.R. and B.S., with input from others. B.S. and P.St.J.R. supervised the project.

**Data availability:** The data that support the plots within this paper and other findings of this study are available from the corresponding authors upon reasonable request.

**Additional information:** Supplementary sections S1–S5

## Methods

**Experimental setup for Stokes wave measurement.** The CW light is amplified in an erbium-doped fibre amplifier (EDFA) and the polarization state and spatial phase are controlled using a combination of circular polarizer (linear polarizer and quarter-wave plate) and Q-plate. A flip-mirror was used to switch the Stokes signal between two paths. In the first path a circular polarizer was used to measure the Stokes polarization state, and a near-field Brillouin scanning analyser (NBA), comprising xy-scanning stage, narrow-band notch filter (6 GHz) and power meter, was used to precisely monitor the Stokes mode profile and power, and to eliminate Fresnel reflections and Rayleigh scattering. In the second path the Stokes signal was split in two, one half was spatially filtered in a single-mode fibre to produce a divergent near-Gaussian beam which was then superimposed on the other half, resulting in spiral patterns of fringes related to the topological charge. These patterns were imaged using a CCD camera after filtering out any stray pump light with a narrow-band filter.

**Near-field Brillouin scanning analyser (NBA).** This system contains an objective lens, an xy-fibre raster scanning stage that is automatically controlled by computer with close-loop feedback, a narrow-band filter (6 GHz) and signal detection equipment (e.g., power meter, optical spectrum analyser, heterodyne system). The Brillouin-shifted signal light, which contains small amounts of pump light (caused by Fresnel reflections or Rayleigh scattering), is collected pixel by pixel by a fibre raster scanning stage. The signal is then filtered to remove the pump light and detected and analysed by above-mentioned detection equipment. The scanning area is $12\times12$ $\mu m^2$ and contains $24\times24$ pixels (each pixel area is 0.25 $\mu m^2$). To increase the resolution of the near-field imaging, we used a highly nonlinear fibre with core diameter 2.4 μm and numerical aperture 0.41 (Nufern UHNA7).

**Experimental setup for reconfigurable vortex isolator.** The CW laser light at 1550 nm is divided at a fibre coupler. One part is amplified using an EDFA and split into signals in the forward and backward directions using a second fibre coupler. The other part is frequency down-shifted or up-shifted using an electro-optic single-sideband modulator (SSBM) and acts as the control wave. The circular polarization states of all the different waves are independently controlled by adjusting the fibre polarization controllers (FPCs) in each optical path and the optical vortices are optionally generated by Q-plates. The setup in the coloured frame in Fig. 5a is the key part of nonreciprocal system: the polarization state in each path can be sequentially converted from circular to linear and from linear to circular using two λ/4 plates, and the control signal can be combined with or separated from the forward and backward signal waves using a polarizing beam-splitter (PBS). After interacting with the control wave, the forward and backward signal waves are extracted using beam splitters (BS) and measured by the NBA systems. The chiral fibre acts as a nonreciprocal isolator when $f_{ctrl} = f_{sig} - f_{SBS}$ and a nonreciprocal amplifier when $f_{ctrl} = f_{sig} - f_{SBS}$.

# Supplementary Information for

# "Nonreciprocal vortex isolator by stimulated Brillouin scattering in chiral photonic crystal fibre"

Xinglin Zeng[1], Philip St.J. Russell[1], Christian Wolff[3], Michael H. Frosz[1], Gordon K. L. Wong[1] and Birgit Stiller[1,2]

[1]Max Planck Institute for the Science of Light, Staudtstr. 2, 91058 Erlangen, Germany

[2]Department of Physics, Friedrich-Alexander University, Staudtstr. 2, 91058 Erlangen, Germany

[3]Center for Nano Optics, University of Southern Denmark, Campusvej 55, DK-5230 Odense M, Denmark

*Corresponding author: xinglin.zeng@mpl.mpg.de

# S1. Heterodyne setup for Brillouin frequency measurement

A special-designed Brillouin heterodyne setup was used to measure the Brillouin frequency in chiral PCF, as shown in Fig. S1. It is based on heterodyne detection in which the backscattered Stokes light from the PCF is coherently coupled with a local oscillator (LO) to get a beat note in the electrical domain. Both pump and LO are derived from a narrow linewidth (<1 kHz) 1550 nm continuous wave (CW) laser using a fibre coupler. The pump wave is then boosted in an Erbium-doped fibre amplifier (EDFA) and injected into the chiral PCF via an optical circulator. The circular polarization states are adjusted by a fibre polarization controller (FPC) placed before a circulator. The vortex generating module (polarizer, λ/4 plate and Q-plate) is optionally used to generate a circularly-polarized vortex-carrying pump signal. The noise-seeded Stokes signal from the PCF is delivered by the circulator and interferes with the LO using the second 90:10 fibre coupler. Narrow-band (6 GHz) notch filters in the path of Stokes signal are used to filter out Fresnel reflections and Rayleigh scattering. The resulting beat-note is detected in the radio-frequency domain with a fast photodiode (PD) and the averaged Brillouin spectra is recorded with an electrical spectrum analyzer.

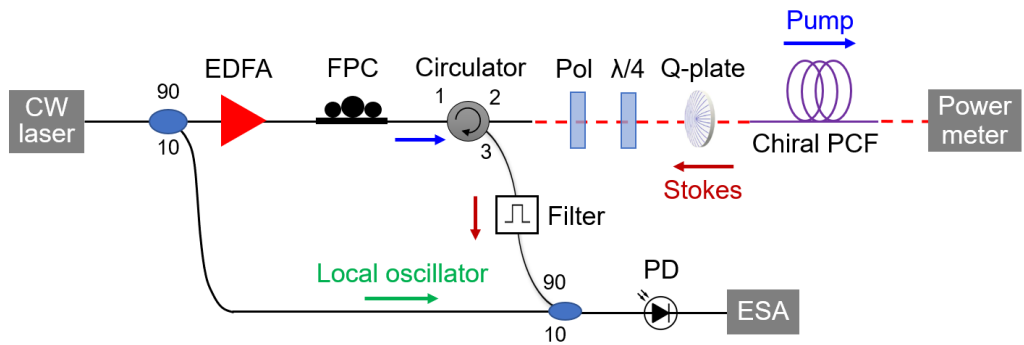

**Fig. S1 | Experimental setup for measuring spontaneous Brillouin spectra in chiral PCF.** A circularly polarized pump signal is generated using a fibre polarization controller (FPC) and a circularly-polarized vortex-carrying pump signal is generated with both a FPC and a vortex generating module. EDFA: Erbium-doped fibre amplifier, Pol: polarizer, λ/4: λ/4 plate, PD: photodetector, ESA: electrical spectrum analyzer.

# S2. SBS measurement for [0, ±1] modes in chiral PCF

In this section, we show Stokes power measurements for a circularly polarized $LP_{01}$-like [0, ±1] pump. Fig. S2a shows the setup. As in all the experiments, the PCF was 200 m long. CW pump light was amplified in an EDFA and its polarization state is controlled using a combination of polarizing beam splitter (PBS) and λ/4 plate. Back-scattered signals with the same polarization state as the pump are transmitted by the PBS and detected by a power meter placed at port 3 of the circulator, while orthogonally polarized light is reflected and detected by a second power meter. The transmitted pump power is monitored by a third power meter. Narrow-band (6 GHz) notch filters in the path of each Stokes signal are used to filter out Fresnel reflections and Rayleigh scattering. Fig. S2b shows the power dependence of the Stokes and transmitted pump signals for $[0, +1]_P$ (LCP), $[0, -1]_P$ (RCP) and $[0, +1]_P$ + $[0, -1]_P$ (linearly polarized light). Above a threshold of ~1.1 W (independent of the pump polarization state), the orthogonally polarized Stokes signal grows rapidly with a slope efficiency of ~60%, while the transmitted pump power saturates. The co-polarized Stokes signal shows no gain, as angular momentum conservation would otherwise be violated. More interestingly, when a linearly polarized pump is injected into the chiral PCF, the polarization state of the backscattered Stokes wave is also linearly polarized and has same azimuthal angle as that of the pump, as shown in the most right one of Fig. S2b. We attribute this to well-controlled optical activity in the chiral PCF, which causes the linearly polarized pump and backward Stokes modes to be co-polarized at all points along the fibre [1].

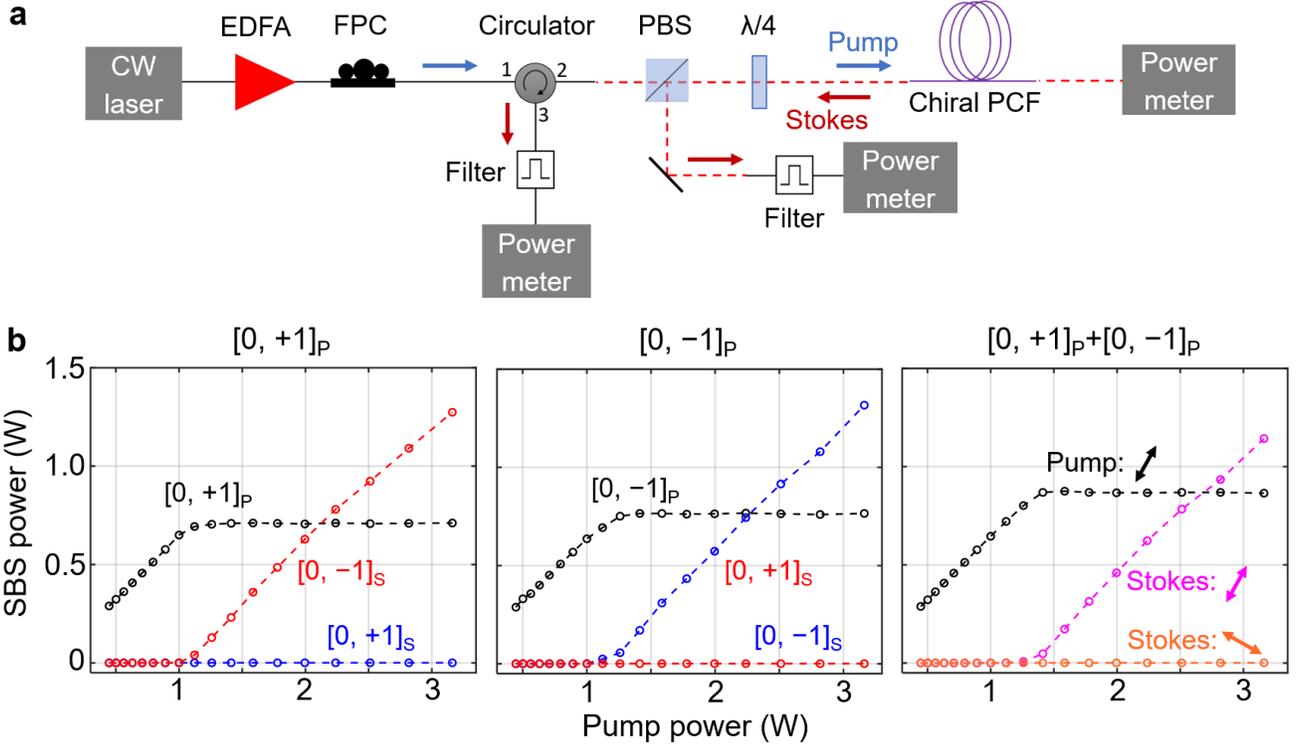

**Fig. S2 | Stokes measurement in PCF $C_3$ with [0, ±1] mode pump. a,** Experimental setup for measuring SBS power (threshold) for circularly polarized pump in chiral PCF. **b,** Stokes and transmitted pump power in a 200 m length of chiral PCF for $[0, +1]_P$, $[0, −1]_P$ and linearly polarized pump light ($[0, +1]_P+[0, −1]_P$).

## S3. Stokes parameters $|S_3|$ measurement for Stokes wave in chiral PCF

To assess the robustness of topology- and polarization-selectivity in chiral SBS, the signal powers and Stokes polarization states were measured for $[\ell_T, +1]$ and $[\ell_T, −1]$ pump light, as shown in Table 1. The fibre was spooled with a diameter of 16 cm in all the measurements. In the measurements, the power of the $[0, ±1]$ and $[±1, ±1]$ pump signals entering PCF $C_3$ was kept at 3.162 W and the power of the $[±2, ±1]$ pump signals entering PCF $C_6$ was 1.26 W. The modulus of the Stokes parameter $S_3$ of backscattered Stokes wave, defined as:

$$|S_3| = \left| \frac{P_{RC} - P_{LC}}{P_{RC} + P_{LC}} \right|, \qquad (2)$$

ranges in value from 0.96 to 0.99 in the experiments, showing that circular polarization states very well preserved.

## S4. Measurement of Brillouin gain coefficient

The Brillouin gain coefficient $g_B$ and spectrum were obtained using a typical pump-seed setup with polarization control, as shown in Fig. S3a. Both pump and seed were derived from a narrow linewidth 1550 nm CW laser, the seed light being frequency tuned using a single side-band modulator (SSBM).

The pump signal was boosted by an EDFA and the polarization states of both pump and seed were controlled using fibre polarization controllers (FPCs). Vortex generating modules (circular polarizer, Q-plate and λ/2 plate) were optionally used to generate circularly-polarized vortex-carrying pump signals. After propagating backwards through the chiral PCF, the seed signal is reflected by a beam splitter (BS) and filtered by a circular polarizer (polarizer and λ/4 plate). Th flip mirror is used to deliver the Stokes light to either a narrow-band filter and power meter for gain coefficient measurement on [0, ±1] modes or an NBA system (for more details see "Methods" in main manuscript) for gain coefficient measurement on [±1, ±1] modes and [±2, ±1] modes. Fig. S3b shows the measured Brillouin gain spectrum for [0, ±1] modes at a pump power of 0.8 W and a Stokes seed power of 10 mW. The peak gain coefficient is 0.169 $W^{-1}m^{-1}$ for [1, +1] pump and the tiny difference in peak gain in the [0, +1] and [0, −1] pump cases is attributed to slight circular dichroism in propagation loss.

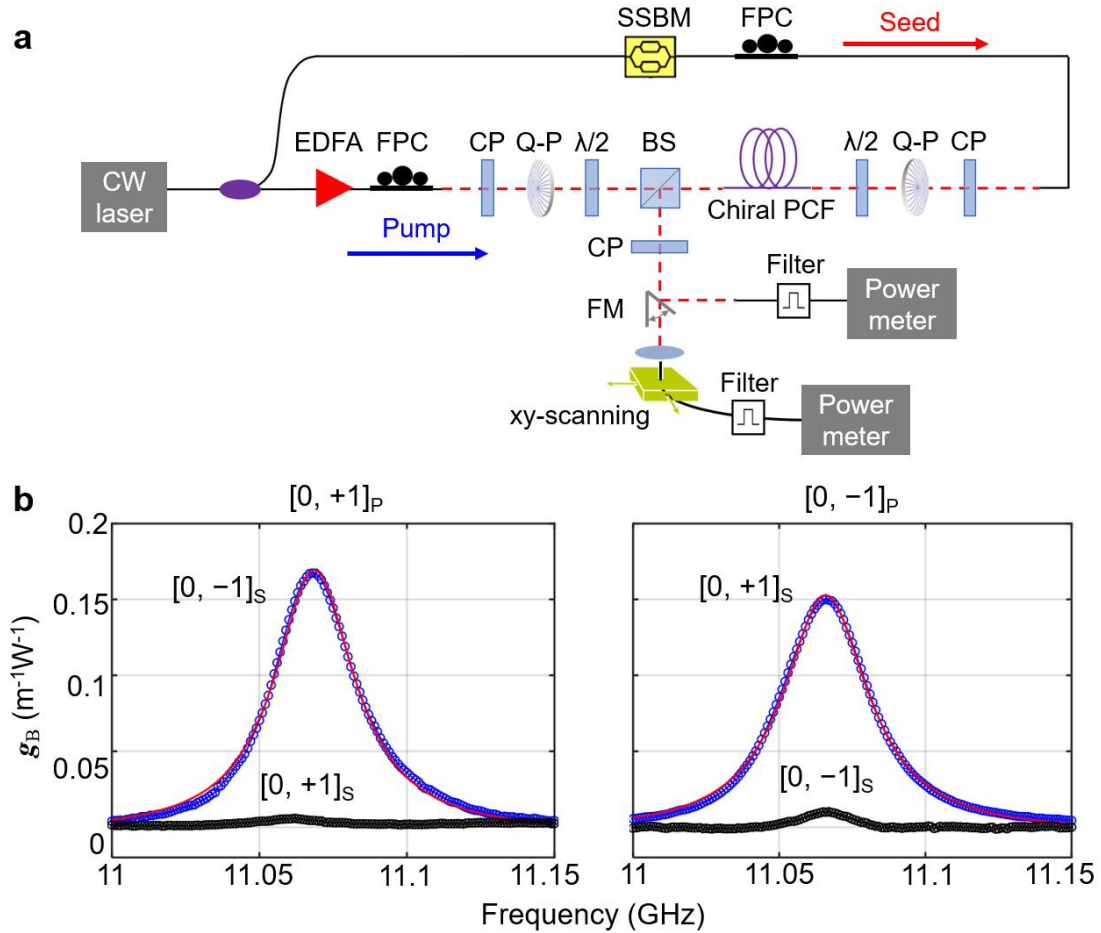

**Fig. S3 | Brillouin gain spectra measurement in chiral PCF. a,** Experimental setup for measuring Brillouin gain spectra of chiral PCF. **b,** The measured Brillouin gain spectra when pumping with [0, +1]$_P$ mode (left) and [0, −1]$_P$ mode (right). The circles are measured data, and the red lines are Lorentzian fittings. FM: flip mirror, CP: circular polarizer, Q-P: Q-plate.

## S5. Measurement of nonreciprocal isolation and amplification for [0, ±1] modes

Fig. S4a shows the isolation of a circularly polarized LP$_{01}$-like signal [0, ±1]$_{sig}$ in the C$_3$ PCF versus control light power. The isolation is as high as 27.5 dB, for signal power 0.617 W and control power 1.78 W. A nonreciprocal circularly-polarized amplifier can also be implemented by switching the control light frequency from $f_0 − f_{SBS}$ to $f_0 + f_{SBS}$. The measured maximum gain is 33 dB, for signal power 0.65 mW and control power 1.78 W. Higher isolation and amplification factors can be achieved for higher control power or longer fibre length. The isolation remains nearly constant (within ~2dB) over a 35 dB dynamic range of signal power for a control power of 1.78 W, as shown in Fig. S4b.

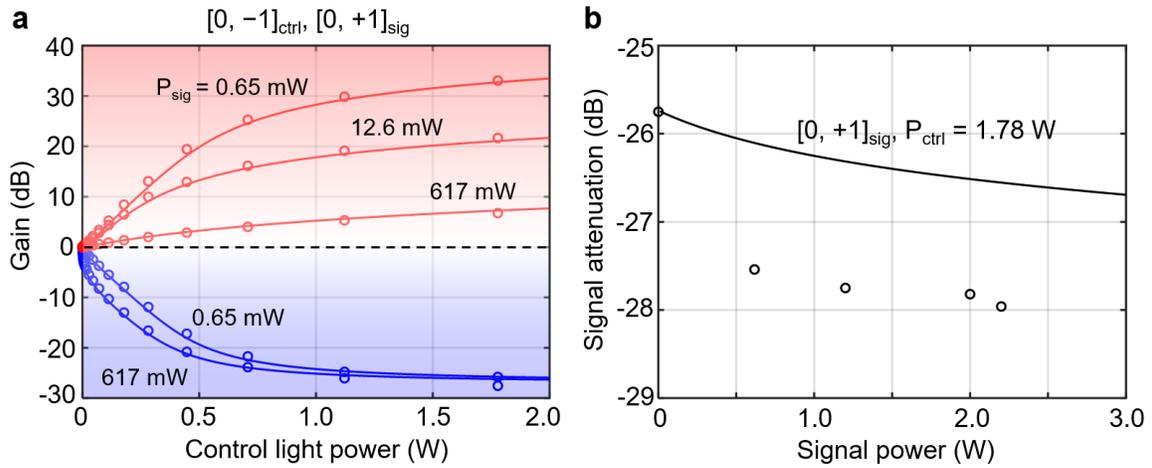

**Fig. S4 | Nonreciprocal attenuation and amplification for [0, ±1] modes. a,** Dependence of attenuation (blue) and amplification (red) of [0, +1] on control light power. **b,** Dependence of signal [0, +1] attenuation on signal power. The circles are experimental datapoints and the curves are theoretical predictions.